\documentclass[10pt,aps,prd,twocolumn,notitlepage,bibnotes,longbibliography,
floatfix,showpacs,citeautoscript,superscriptaddress]{revtex4-1}

\usepackage{graphics}
\usepackage{graphicx}

\usepackage{amsmath}
\usepackage{amssymb}
\usepackage{mathrsfs}
\usepackage{bbm}
\usepackage{braket}

\newcommand{\la}[1]{\label{#1}}

\usepackage[pdfstartview=FitV, pdfusetitle,
			bookmarks=true,bookmarksnumbered=false,bookmarksopen=false,
			breaklinks=true,pdfborder={0 0 0},backref=false,colorlinks=true, 
			linkcolor=blue, citecolor=blue, urlcolor=blue]{hyperref}
\usepackage{orcidlink}

\begin{document}

\title{Geometric Qubits in Programmable Atomic Trimers}

\author{Julien Garaud\,\orcidlink{0000-0001-5087-3115}}
\email{garaud.phys@gmail.com}
\affiliation{Universit\'e de Tours, Universit\'e d’Orl\'eans, 
			CNRS, IDP, UMR 7013, Tours, France}
\author{A.J. Niemi,\orcidlink{0000-0003-3408-5834}}
\email{Antti.Niemi@su.se}
\affiliation{Nordita, Stockholm University, Roslagstullsbacken 23, 
			SE-106 91 Stockholm, Sweden}
\affiliation{Wilczek Quantum Center,  Shanghai Institute for Advanced Studies, 
			University of Science and Technology of China, Shanghai 201315, China}

\begin{abstract}
Motivated by programmable tweezer arrays, we develop an exactly solvable shape-space 
theory for near-equilateral atomic trimers. The Higgs oscillator on Kendall's shape 
sphere gives a nonresonant spectral lattice whose fixed-angular-momentum 
relative-stationary minimizers are supported on two adjacent admissible sites. 
These symmetry-selected doublets are separated from higher relative-stationary branches 
by a finite branch gap. With phase-coherent shape-mode driving and Rydberg-mediated 
conditional phases, they furnish a candidate route to geometric qubits and entangling 
operations.
\end{abstract}

\maketitle

\paragraph*{Introduction:} Optical-tweezer arrays of individually trapped neutral 
atoms such as {\tt Rb}, {\tt Cs}, or {\tt Sr} provide a versatile platform for assembling 
atomic trimers with Rydberg-mediated interactions\cite{Nogrette-2014,Barredo-2014,
Labuhn-2016,Barredo-2016,Browaeys-2020,Chew-2024}. Beyond single-atom encodings, 
both the trimer geometry and internal couplings can be tuned dynamically via tweezer 
rearrangement, optical detunings, and Rydberg dressing \cite{Saffman-2010,Pupillo-2010,
Johnson-2010,Lorenz-2021,Steinert-2023}. This motivates engineered trimer arrays in which 
qubits are encoded in collective shape modes and steered along programmable paths in the 
triangle's shape space. The associated shape-space symmetries can protect collective 
qubit subspaces, organize their phase winding, and constrain leakage from the encoded 
manifold.

Here we formulate a symmetry-based effective theory for such atomic trimers and trimer 
arrays in programmable, site-addressable neutral-atom tweezers. We focus on collective 
shape dynamics rather than microscopic short-range details. We assume a separation of 
scales in which the center of mass is pinned, out-of-plane motion is suppressed, 
and overall scale fluctuations are frozen or decoupled by the external confinement. 
This isolates shape as the collective coordinate most relevant for spectroscopy and 
gate design. Switchable couplings between trimers can be generated by transient access 
to Rydberg states, either through short resonant pulses \cite{Saffman-2010,Chew-2022,
Evered-2023} or through off-resonant dressing in tweezer arrays \cite{Pupillo-2010,
Johnson-2010,Lorenz-2021,Steinert-2023}. In this regime, trimer assemblies furnish a 
tunable setting for effective many-body Hamiltonians with molecule-like collective degrees 
of freedom within ultracold Rydberg physics \cite{Shaffer-2018,Fey-2019}.
 
\paragraph*{Model:}  Kendall's shape-space construction \cite{Kendall-1984,Montgomery-2015} 
represents oriented planar trimer shapes by the Kendall two-sphere $\mathbb S^2_K$. 
The poles are the two oppositely oriented equilateral configurations and the equator 
is the collinear locus. For a trimer in three-dimensional space the two poles are related 
by a global $\rm{SO}(3)$ rotation. After  quotienting by overall rotations, the physical 
unoriented shape space is then represented by a closed hemisphere of $\mathbb S^2_K$,
bounded by the collinear locus  as shown in figure \ref{fig-1}.
%
\begin{figure}[!htb]
\hbox to \linewidth{ \hss
\includegraphics[width=0.925\linewidth]{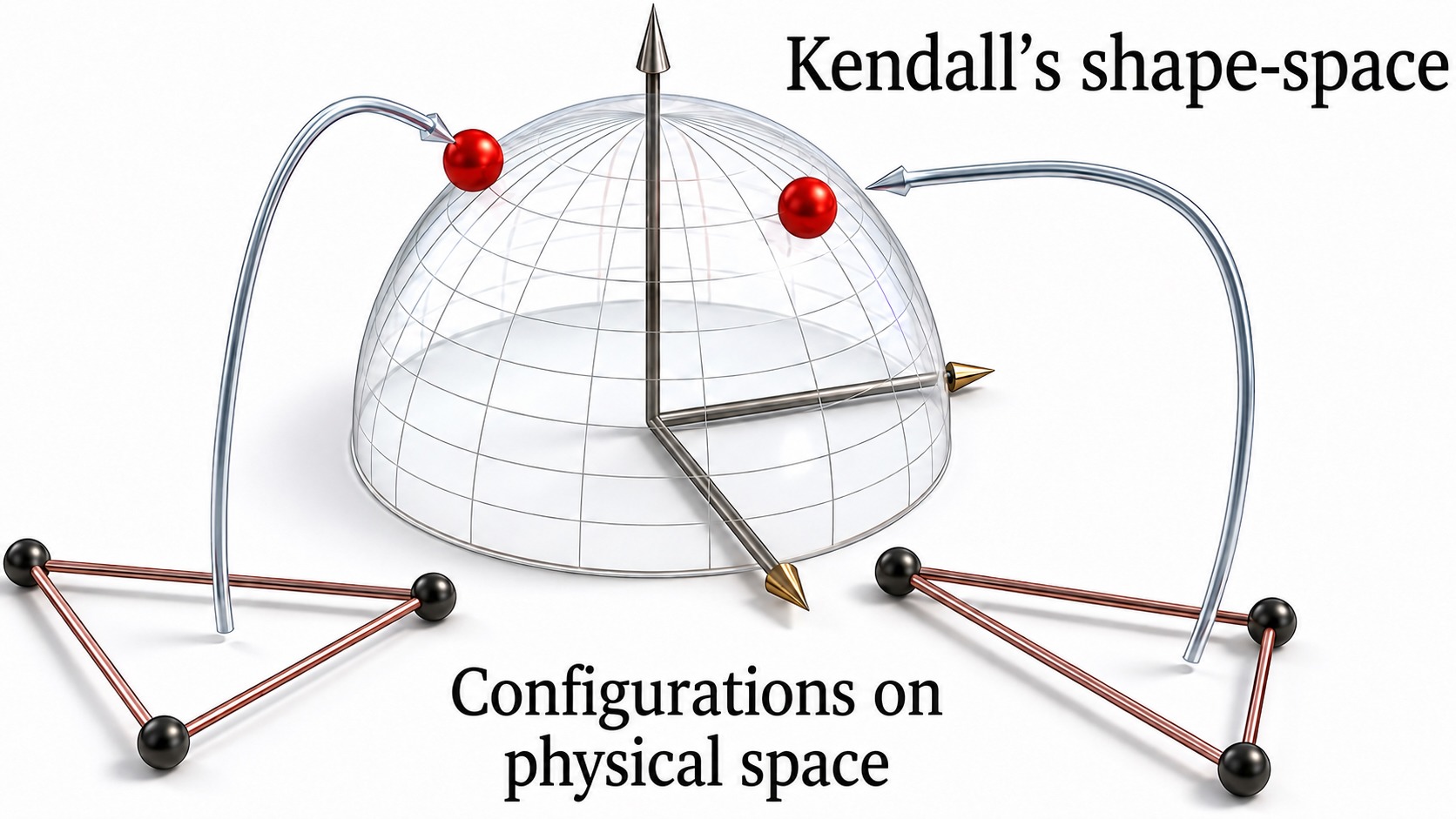}
\hss}
\caption{Kendall's space of triangular shapes is a hemisphere, with points identified 
with a different shape of a triangle up to an overall scale and spatial orientation.
}
\la{fig-1}
\end{figure}

We model dynamics on this hemisphere by the Higgs oscillator. Near the equilateral pole 
it reproduces the doubly degenerate harmonic $E$-mode, while it globally gives an exactly 
solvable spherical model with an effective barrier at collinearity.
Remarkably, its noncommensurate energy spectrum avoids the resonant ladder structure 
of a harmonic oscillator, thereby helping to isolate two-state spectral supports 
as natural encoded doublets.
Phase-programmable Raman driving can then address the low-lying shape states 
\cite{Allen-1992,Andersen-2006,Lunt-2024}, enabling geometric qubit operations along 
controlled paths on the shape sphere \cite{Berry-1984,WilczekZee-1984,
Zanardi-1999,Sjoqvist-2012,Xu-2012,Zhang-2023,Dai-2025}.

To formulate the shape dynamics, we  describe a trimer by two mass-scaled Jacobi vectors
$\boldsymbol{\xi}_1,\boldsymbol{\xi}_2$ \cite{Blume-2012}. The moment ofinertia is 
$I=|\boldsymbol{\xi}_1|^2+|\boldsymbol{\xi}_2|^2$, so fixing the overall scale restricts 
the dynamics to the preshape space $I=I_0\simeq \mathbb S^3$. 
Simultaneous planar rotations of $\boldsymbol{\xi}_1$ and $\boldsymbol{\xi}_2$ define 
the Hopf fibration and quotienting by this \, $\mathbb S^1$ action gives Kendall's 
shape sphere $\mathbb S^3/\mathbb S^1\cong\mathbb S^2_K$ \cite{Kendall-1984,Montgomery-2015}. 
We use a stereographic complex coordinate $z\in\mathbb C$ on $\mathbb S^2_K$, defined 
from the south pole so that the chart contains the north pole and $z=0$ represents 
the chosen equilateral minimum. For an array of $\mathcal N$ trimers, the internal 
configuration space is $(\mathbb S^2_K)^{\mathcal N}$ and the effective low-energy 
Hamiltonian $\hat{\mathrm H}$ is assumed additive up to controllable, shape-dependent 
interactions.

As the spherical analogue of the isotropic harmonic oscillator, the Higgs oscillator
\cite{Higgs-1979,Leemon-1979,Supplementary-arxiv} provides a natural exactly solvable 
low-energy Hamiltonian for engineered shape dynamics. With the equator  $|z|=1$ 
representing the collinear-triangle locus, and setting $\hbar=1$ with energies measured 
in oscillator units and time in units of the inverse characteristic shape frequency,
the Hamiltonian is
\begin{equation}
\hat{\mathrm H} =
- 2(1+z\bar z)^2\,\partial_z\partial_{\bar z}
+\frac{1}{2} \frac{z \bar z}{ (1-z\bar z)^2} \,.
\la{Ham1}
\end{equation}
The first term is the Laplace operator on the  Kendall sphere expressed in stereographic 
coordinates. Near the equilateral point the model produces  the universal harmonic limit 
of any rotationally invariant stable potential, and the potential diverges at $|z|=1$,  
thereby energetically excluding the collinear locus. 
The full Hamiltonian is invariant under $z\mapsto 1/\bar z$, which exchanges the
two hemispheres of $\mathbb S^2_K$. For a freely embedded three-dimensional
trimer, the two oriented planar representatives related by this map are connected
by a proper global $SO(3)$ rotation in ambient space. The Hamiltonian is furthermore 
invariant under shape rotations about the equilateral point $z=0$, generated by 
\begin{equation*}
\hat{\mathrm J}=z\,\partial_z-\bar z\,\partial_{\bar z}\,,  \ \ \ {\rm and}  \ \ \   
[\hat{\mathrm H} , \hat{\mathrm J} ] = 0\,.
\end{equation*}
In analogy with generalized coherent states \cite{Perelomov-1986} we seek normalized 
solutions $\ket{\psi (t)}$ of the time-dependent Schr\"odinger equation whose evolution 
closes on the shape space symmetry orbit generated by $\hat{\mathrm J}$. The reference 
state $\ket{\psi}\equiv \ket{\psi(0)}$ is determined dynamically as a critical point of 
$ \langle\hat{\mathrm H}\rangle_\psi\equiv \bra{\psi}\hat{\mathrm H}\ket{\psi}$ at fixed 
norm $\langle \psi | \psi \rangle =1$ and fixed angular-momentum expectation value 
$\langle\hat{\mathrm J}\rangle_\psi=\bra{\psi}\hat{\mathrm J}\ket{\psi}   =\ell $.
For such coherent  {\it relatively stationary} states 
\footnote{
Relative stationarity generalizes ordinary stationarity in systems with continuous
symmetries: the state is stationary only after quotienting by the symmetry flow.
The classical analogues are relative equilibria, such as steadily rotating rigid
bodies, rotating vortex patterns, or travelling waves in a co-moving frame.
} 
there exist real parameters $\lambda_N$ and $\lambda_J$ such that
\begin{equation}
\ket{\psi(t)} = e^{-i\hat{\mathrm H} t} \ket{\psi} \ = \  
e^{-i\lambda_N t}e^{-i\lambda_J\hat{\mathrm J}t}\ket{\psi}  
\la{Hevo1}
\end{equation}
so the Hamiltonian flow preserves the corresponding group orbit. For $\lambda_J=0$ 
one recovers ordinary stationary states, with $\lambda_N$ the energy eigenvalue of 
$\hat{\mathrm H}$. But for generic $\ell$ the state is not an energy eigenstate, 
and without loss of generality here we focus on the positive-$\ell$ sector.

\paragraph*{Relative stationary states:}  
The energy eigenvalues can be labeled by the radial quantum number $n_r\in\mathbb Z_{\ge0}$ 
and by the angular-momentum quantum number  $m\in\mathbb Z$. With $n=2n_r+|m|  $, 
where $n_r\in\mathbb Z_{\ge 0} $, $n \geq |m| $ and $n-m \in 2\mathbb Z$ the energy 
eigenvalues are \cite{Higgs-1979,Leemon-1979,Supplementary-arxiv}
\begin{equation}
E_{nm}\equiv E_n =
2(n+1)^2+\sqrt{5}\,(n+1)
\la{Enm}
\end{equation}
so that the energy of an eigenstate $\ket{n,m}$ of $\hat{\mathrm H}$ depends only 
on the shell index $n$, with  $m$ distinguishing the states within a given shell: 
The shell degeneracy reflects  a dynamically generated $\rm{SU}(2)$ symmetry of the 
Higgs-oscillator spectral lattice, with each fixed-$n$ shell forming an irreducible 
multiplet.  

We expand the normalized time dependent wavefunction \eqref{Hevo1} in terms of the 
eigenstates $\ket{n,m}$,
\begin{equation*}
\ket{\psi(t)} =\sum_{n,m}  c_{nm}(t)\,\ket{n,m} 
\ \  \& \\ \ 
1=\langle \psi | \psi \rangle =\sum_{n,m} p_{nm},
\end{equation*}
where the probabilities $p_{nm}\equiv |c_{nm}|^2 \in [0,1]$ are time independent under 
Hamiltonian evolution. Furthermore, the average energy and shape-space angular momentum 
are respectively
\[
E =\langle\hat{\mathrm H}\rangle_\psi =\sum_{n,m}  E_n p_{nm} \ \ \&  \  \  
\ell =\langle\hat{\mathrm J}\rangle_\psi=\sum_{n,m} m p_{nm}, 
\]
and from Eq.\eqref{Hevo1} we get
\[
\bigl[E_{n}-\lambda_N-\lambda_J m\bigr]\,c_{nm}(t)=0 \ \ \ \ \  \forall\,n,m \,,
\]
so that every occupied mode $c_{nm}(t)\not= 0  $ must satisfy
\begin{equation}
E_n\equiv 2(n+1)^2+\sqrt5\,(n+1) = \lambda_N+\lambda_J m\,.
\la{E-spec}
\end{equation}
Since $E_n$ is quadratic in $n$, this condition defines a parabola on the $(n,m)$ lattice. 
For any admissible pair of states $(n_a,m_a)$ and $(n_b,m_b)$ with $n_a\neq  n_b$ and 
$m_a\neq m_b$, the corresponding support parabola is fixed by
\begin{equation}
\lambda_N =\frac{m_aE_{n_b}-m_bE_{n_a}}{m_a-m_b} 
 \ \ \ \ \& \ \ \ \   
\lambda_J=\frac{E_{n_a}-E_{n_b}}{m_a-m_b}  \,.
\la{lambdas}
\end{equation}
As shown in the Supplementary Material \cite{Supplementary-arxiv}, for $ \lambda_J\neq 0$  
the support parabola through two distinct lattice points $(n_a,m_a)$ and $(n_b,m_b)$ 
contains no further admissible lattice points. Therefore the relative-stationarity 
condition selects only the two chosen modes, making relative stationary qubit encodings 
generic in the Higgs-oscillator formulation. This is illustrated in Figure~\ref{fig-2}.

\paragraph{Energy minimizer:} 
Let $k\in\mathbb Z_{\geq 0}$ be such that $ k<\ell<k+1 $. Among all normalized states 
with fixed $\ell$, the minimum of $\langle\hat{\mathrm H}\rangle_\psi$ is attained by 
the family of relatively stationary qubits  supported on the two adjacent extremal 
states $\ket{k,k}$ and $\ket{k+1,k+1}$. The  normalized time-dependent minimal energy 
superposition is 
\begin{align}
 & \ket{\psi_{k\ell}(t)}  =
\la{miniE}   \\
&\, e^{-iE_k t} \! \left[  \sqrt{p^{(k)}_0}\,\ket{k,k} +
e^{ i(\varphi_k- \Delta_k t) } \sqrt{p^{(k)}_1}\,\ket{k+1,k+1} \right]
\notag
\end{align}
where $p^{(k)}_0=k+1-\ell $ and $p^{(k)}_1=\ell-k=1-p^{(k)}_0$ and 
$\Delta_k\equiv E_{k+1}-E_k=4k+6+\sqrt5$. To show this,  we conclude from \eqref{E-spec} 
and \eqref{lambdas} that for every admissible lattice point 
\begin{align*}
& E_n-\lambda_N^{(k)}-\lambda_J^{(k)}m  = \notag \\
& ~ 2(n-k)(n-k-1) +(4k+6+\sqrt5)(n-m) \ge 0  ,
\end{align*}
with equality only at $(n,m)=(k,k)$ and $(k+1,k+1)$. Hence every normalized
state with $\langle\hat{\mathrm J}\rangle_\psi=\ell$ satisfies
$\langle\hat{\mathrm H}\rangle_\psi \ge \lambda_N^{(k)}+\lambda_J^{(k)}\ell $,
and this bound is saturated only by states supported on $\ket{k,k}$ and $\ket{k+1,k+1}$.

\begin{figure*}[!htb]
\hbox to \linewidth{ \hss
\includegraphics[width=0.925\linewidth]{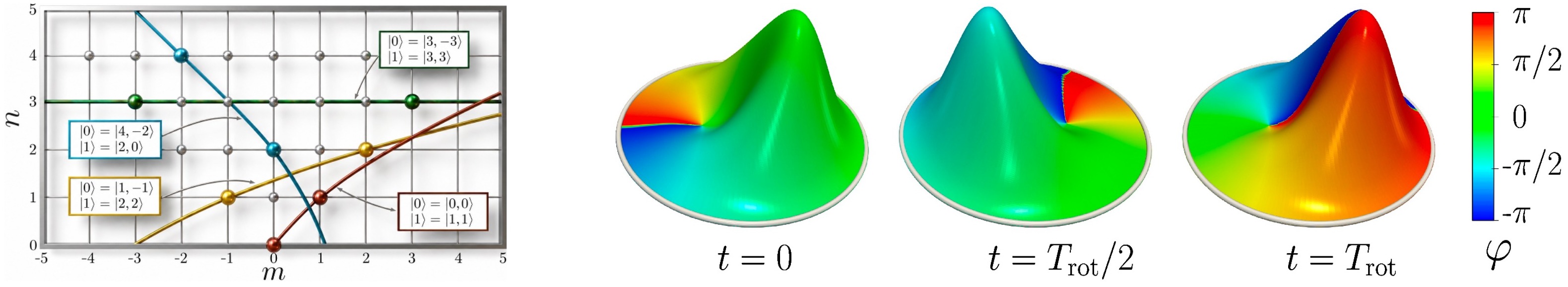}
\hss}
\caption{The leftmost panel shows Higgs-oscillator lattice states $\ket{n,m}$. 
Nontrivial relatively stationary states are supported by two lattice points on a support 
parabola \eqref{E-spec}. The red parabola, $\ket{0,0}\leftrightarrow\ket{1,1}$, 
is the energy minimizer for $0<\ell<1$; the yellow curve shows an opposite-chirality 
minimizer for $-1<\ell<2$; the blue curve illustrates negative $\ell$, and the green line 
a $\lambda_J=0$ degeneracy. The remaining panels display the time evolution of the $k=0$, 
$\ell=3/4$ minimizer \eqref{miniE}, where the elevation denotes the density $|\psi|$ and 
the colormap, the phase $\arg\psi$. The modulus and phase have different periods implying 
that the time evolution accumulates a geometric phase \eqref{geomp}; see Supplemental 
Material for animation \cite{Supplementary-arxiv}. 
}
\la{fig-2}
\end{figure*}

\paragraph*{Qubit:} 
We identify $\ket{k,k} = \ket{0}_k$ and $\ket{k+1,k+1} = \ket{1}_k$ as the logical basis. 
Their probabilities can be expressed in terms of the Bloch-sphere polar angle $\theta_k$ 
as $p^{(k)}_0 = 1-p^{(k)}_1 =\cos^2\frac{\theta_k}{2}$, and their initial relative phase 
$\varphi_k$ is the azimuthal angle, freely evolving according to 
\[
\varphi_k(t)=\varphi_k - \Delta_k t \,,\quad\text{with}\quad
\Delta_k -(4k+6+\sqrt5)t \,.
\]
The density rotates rigidly in shape space with the period
$T_{\rm rot}^{(k)}= {2\pi}/{\Delta_k}$,
and the state also acquires the geometric phase \cite{Supplementary-arxiv}
\begin{equation}
\beta^{(k)}_{\rm geom} =
\pi(1-\cos\theta_k) =
2\pi(\ell-k) 
\quad (\mathrm{mod}\ 2\pi)\,.
\la{geomp}
\end{equation}
The rotation of the density defines an internal shape-space clock:
a two-state superposition prepares a Ramsey protocol, the relative
phase accumulates during free evolution, with the resulting longitude converted 
into a population imbalance for readout.
See Figure~\ref{fig-2} and animation in \cite{Supplementary-arxiv}.

\paragraph*{Excited states:} For the lowest branch
$0<\ell<1$ the next-lowest relatively stationary two-state branch changes at
$\ell_c\equiv (6+\sqrt5)/(8+\sqrt5)$. For $0<\ell<\ell_c$ it is supported on $\ket{0,0}$ 
and $\ket{2,2}$, whereas for $\ell_c<\ell<1$ it is the same-shell pair $\ket{1,-1}$ and 
$\ket{1,1}$. The corresponding gap above the minimum-energy qubit is
\begin{equation}
\Delta E_{\rm next}^{(0)}(\ell) =
\min\Bigl\{\,2\ell,\,(6+\sqrt5)(1-\ell)\Bigr\}\,.
\la{E-gap}
\end{equation}
At $\ell=\ell_c$ the two branches are degenerate. The same-shell branch has
$\lambda_J=0$ and is therefore an ordinary energy eigenstate.

For $k\geq1$ and $k<\ell<k+1$, let $x=\ell-k$. The first subleading
positive-chirality branches are 
\[
\ket{k,k}\leftrightarrow\ket{k+2,k+2},
\ 
\ket{k-1,k-1}\leftrightarrow\ket{k+1,k+1}.
\]
Their gaps above the minimizing branch $\ket{k,k}\leftrightarrow\ket{k+1,k+1}$ are 
respectively $\Delta E_R^{(k)}=2x $ and $\Delta E_L^{(k)}=2(1-x)$. Thus the next-lowest 
branch is the first of these for $k<\ell<k+\tfrac12$, and the second for 
$k+\tfrac12<\ell<k+1$; at $\ell=k+\tfrac12$ they are degenerate. Hence
\[
\Delta E_{\rm next}^{(k)}(\ell)
=
2\min\{x,1-x\}, \qquad k\geq1 .
\]

Branches with both chiralities also enter the relatively stationary spectrum.
At fixed $\ell$, their lowest-energy representatives are obtained from
pairs
\[
\ket{|m_-|,m_-}\leftrightarrow\ket{m_+,m_+},
\quad\text{with}\quad
m_-<0<m_+ \,,
\]
since $E_n$ depends only on $n$ and admissibility requires $n\geq |m|$. For $k\geq1$, 
the lowest opposite-chirality branch is $\ket{1,-1}\leftrightarrow\ket{k+1,k+1}$ with gap
\[
\Delta E_{\rm opp}^{(k)}(\ell)
=
\frac{2(1-x)}{k+2}\left(k^2+3k+6+\sqrt5\right)\,.
\]
This lies above the relevant first subleading positive-chirality branch throughout 
the open interval $k<\ell<k+1$; the case $k=1$ is shown in Figure~\ref{fig-2}.

Finally, same-shell stationary states do not change this ordering for $k\geq1$. 
The lowest $\lambda_J=0$ manifold capable of realizing $k<\ell<k+1$ is the shell $n=k+1$, 
with energy $E_{k+1}$. Its gap above the minimum branch is
\begin{align}
\Delta E_{\rm shell}^{(k)}(\ell)
&=
E_{k+1}-E_{\min}^{(k)}(\ell)
= E_{k+1} -  p^{(k)}_0E_k+p^{(k)}_1E_{k+1} \nonumber
\\ 
&= (k+1-\ell)(4k+6+\sqrt5)
\end{align}    
This is also larger than the relevant first subleading nonzero-$\lambda_J$
branch throughout $k<\ell<k+1$. Thus,  for $k\geq1$, the next relatively
stationary qubit is always one of the two nearest positive-chirality branches
above, while opposite-chirality and same-shell branches appear only higher in
the relative spectrum. The interval $0<\ell<1$ is exceptional because the
absence of a left positive-chirality branch allows the same-shell pair
$\ket{1,-1}\leftrightarrow\ket{1,1}$ to be the next-lowest branch near
$\ell=1$.

\paragraph*{On control:} 
To operate on the encoded doublet, a phase-coherent drive must selectively couple 
the two support states. We illustrate this for the lowest branch, $k=0$ and $0<\ell<1$, 
with ${\cal H}_L={\rm Span}\{\ket{0,0},\ket{1,1}\}$ the logical subspace.  
In the stereographic  coordinate $z$, a natural pair of minimal symmetry-allowed 
shape-mode coupling operators is
\[
\hat Q_+=\frac{2z}{ 1+\bar z z},
~~
\hat Q_-=\frac{2\bar z}{ 1+\bar z z},
~~\text{and}~~
[\hat{\mathrm J},\hat Q_\pm]=\pm \hat Q_\pm  \,,
\]
and a minimal phase-coherent monochromatic drive with frequency $\omega_d$ and  Raman-like
selection rules is
\[
\hat {\mathrm V}_{\rm R}(t)
=
({\cal E}/2)e^{-i(\omega_d t+\chi_0)}\hat {Q}_+
+
({\cal E}/2)e^{i(\omega_d t+\chi_0)}\hat Q_- .
\]
It addresses the encoded transition $\ket{0,0}\leftrightarrow\ket{1,1}$ resonantly. 
Projecting onto ${\cal H}_L$ and using the rotating-wave approximation gives the effective 
Hamiltonian
\begin{align*}
  \hat{\mathrm H}_{\rm eff}
&=
 -\frac{\delta}{ 2}\sigma_z + \frac{\Omega}{ 2}
\left(\cos\phi_d\,\sigma_x+\sin\phi_d\,\sigma_y
\right) \,,~~\text{with} \\ 
\delta&=  \omega_d-\Delta_0  \ \ \text{and } \ \ 
\Omega e^{-i\phi_d} = {\cal E}e^{-i\chi_0}\bra{1,1}\hat Q_+\ket{0,0},
\end{align*}
and $\Delta_0=E_1-E_0=6+\sqrt5$. The pulse $\int \Omega(t)\,dt$ controls the population 
transfer and therefore the latitude $\ell$, while the optical phase $\phi_d$ fixes 
the azimuthal direction of the transverse drive and hence controls the relative phase 
$\varphi$. Detuning or free precession supplies additional rotations about the Bloch 
$z$-axis. In Bloch-vector form, with $\mathbf s=\langle\boldsymbol\sigma\rangle$, 
the projected dynamics is
\begin{equation*}
\dot{\mathbf s} =
\left(
\Omega\cos\phi_d,\,
\Omega\sin\phi_d,\,
-\delta \right) \times \mathbf s  .
\end{equation*}
This effective two-level description is valid when $|\Omega|$, $|\delta|$, and the drive 
modulation rate remain small compared with the detuning to allowed off-resonant transitions 
outside ${\cal H}_L$.

\paragraph*{ On arrays:}
Finally we indicate how the single-trimer encoding extends to arrays.  For this, 
we consider two minimum-energy encoded trimers with $k_a<\ell_a<k_a+1$ ($a=1,2$), 
with $\ket{0}_a=\ket{k_a,k_a}$ and $\ket{1}_a=\ket{k_a+1,k_a+1}$ the logical basis.  
The free evolution produces only local phase rotations, with splittings 
$\Delta_a=E_{k_a+1}-E_{k_a}$, which are removable by single-qubit gates.  
Entanglement must therefore come from a genuinely joint operation. A natural one is 
a linked two-trimer holonomic cycle as in the Wilczek--Zee/Chern--Simons construction 
of Ref.~\cite{Dai-2025}, where closed shape cycles have restricted holonomy group 
${\rm SU}(2)$ and linked cycles generate controlled phases. Let $\Gamma_a$ be the closed 
shape-space`control loop of trimer $a$ and $\widetilde\Gamma_a$ its lifted Wilson loop 
in the ambient (physical) three-dimensional space, with relative signed linking number
$L_{12}={\rm Lk}(\widetilde\Gamma_1,\widetilde\Gamma_2)$.  In the diagonal Cartan sector, 
suppose that the logical state $\ket{\mu}_a$, $\mu=0,1$, carries geometric weight 
$q_{a,\mu}$. Up to local dynamical, geometric, and self-linking phases, the multi-trace 
Wilson loop \cite{Dai-2025} gives $\Phi_{\mu\nu}=(4\pi/\kappa)q_{1,\mu}q_{2,\nu}L_{12}$, 
where $\kappa$ is the effective Chern--Simons level. Therefore, the invariant two-qubit 
phase is the linked-cycle Wilson-loop invariant projected onto the encoded shape-qubit 
subspace:
\begin{align*}
\Phi_{\rm ent}
 &= 
\Phi_{11\! }-\Phi_{10}\! -\Phi_{01}\! +\Phi_{00}
\\   &=  
\frac{4\pi}{\kappa}
(q_{1,1\! }-q_{1,0})(q_{2,1}\! -q_{2,0})L_{12}\,.
\end{align*}
After local phase corrections, the projected gate is locally equivalent to
$U_{\rm CP}(\Phi_{\rm ent})={\rm diag}(1,1,1,e^{i\Phi_{\rm ent}})$ in the logical basis 
$\ket{\mu}_a$. Thus the entangling phase is a nonlocal geometric invariant of linked 
shape cycles, not an independently assumed controlled interaction. 
For the simple choice $q_{a,0}=0$, $q_{a,1}=q_a$, and $L_{12}=1$, the linked cycle gives 
$\Phi_{\rm ent}=4\pi q_1q_2/\kappa$.  Tuning this phase to $\pi \, ( {\rm mod}\ 2\pi)$ 
yields $U_{\rm CZ}={\rm diag}(1,1,1,-1)$. Together with a holonomic Hadamard-type operation 
$H_{\rm had}^{(2)}$ on the second encoded trimer, this gives
\begin{equation*}
U_{\rm CNOT}
=
\left(\mathbbm 1^{(1)}\otimes H_{\rm had}^{(2)}\right)\,
U_{\rm CZ}\,
\left(\mathbbm 1^{(1)}\otimes H_{\rm had}^{(2)\dagger}\right) \,.
\end{equation*}
In this way  the linked-cycle phase supplies the entangling primitive, while local
shape-space holonomies and Raman control provide the encoded single-qubit shape-space
rotations.

\paragraph*{Summary and outlook:} 
We have proposed an effective theory of programmable atomic trimers in which quantum 
information is encoded in collective shape degrees of freedom. In the Higgs-oscillator 
description, relative-stationary states are organized by support curves in the spectral 
lattice, and two-point minimum-energy supports define natural encoded qubits protected 
by a finite leakage gap. The control construction shows how this spectral gap can be
used operationally,  with weak phase-coherent drives generating encoded single-qubit
rotations, and linked Wilczek--Zee cycles generating controlled phases in
trimer arrays.

Here we have specified the control matrix elements, the effective Cartan weights,
and the inter-trimer couplings at the effective-theory level. A microscopic
treatment of realistic tweezer potentials must include anharmonic dressing,
finite-size effects, and dynamical leakage during driven cycles. Determining the
range of drive strengths, detunings, and cycle times compatible with low leakage
will decide whether shape-encoded qubits can be implemented as a scalable
neutral-atom architecture.

\begin{acknowledgments}
J.G.  thanks M. Chernodub, 
and  A.J.N.  thanks X.-C. Yao for discussions. A.J.N. is
supported by the Swedish Research Council under Contract No. 2022-04037 
and by COST Action CA23134
(POLYTOPO). 
\end{acknowledgments}

%

\setcounter{equation}{0}
\setcounter{figure}{0}
\setcounter{table}{0}
\setcounter{section}{0}
\makeatletter
\renewcommand{\theequation}{S\arabic{equation}}
\renewcommand{\thefigure}{S\arabic{figure}}
\renewcommand{\bibnumfmt}[1]{[S#1]}
\renewcommand{\citenumfont}[1]{S#1}
\onecolumngrid
\pagebreak
\begin{center}
\textbf{\large Supplementary Material: Geometric Qubits in Programmable Atomic Trimers}
\vskip 0.4cm
\begin{minipage}{0.9\textwidth}\parindent11.10839pt 
\indent 
In the Supplemental Material, we present several additional details, including derivation of the model and details about Higgs oscillator and elements of relative stationary states. We also discuss additional animations.

\end{minipage}
\vskip 0.4cm
\end{center}
\twocolumngrid

\subsection{Higgs oscillator on the stereographic shape sphere}
\label{app:Higgs}

In terms of the stereographic coordinate $z\in\mathbb C$ the round metric and the Laplace-Beltrami on the Kendall shape sphere $\mathbb S^2_K$ read as
\[
ds^2=\frac{dz\,d\bar z}{(1+|z|^2 )^2}   \ \ \text{and} \ \ 
\Delta_g=4(1+|z|^2)^2\,\partial_z\partial_{\bar z}\,.
\]
There, a  quantum Hamiltonian with potential $V(z,\bar z)$ has the general form
\[
\hat {\mathrm H}=-\frac12\Delta_g+V(z,\bar z)
=
-2(1+|z|^2)^2\,\partial_z\partial_{\bar z}+V(z,\bar z).
\]
We   impose the discrete inversion symmetry 
\begin{equation}
{\cal I}:\quad z\longmapsto \frac{1}{\bar z}\,,
\label{inv}
\end{equation}
that changes the orientation of a triangle.
A convenient $\cal I$-invariant basis of real functions 
with the equatorial barrier are
\begin{equation}
X_n(|z|):=\frac{|z|^{2n}}{(1-|z|^2)^2(1+|z|^2)^{2n-2}},
\qquad n\ge1 \,.
\label{radial-basis}
\end{equation}
The first basis function $X_1(|z|)$ is harmonic, while those $X_n(|z|)$ with $n>1$ are anharmonic.

Similarly, the anisotropic  
\begin{equation}
X:=\frac{z+\bar z}{1+|z|^2} \,,\quad\text{and}\quad
Y:=\frac{z-\bar z}{i(1+|z|^2)}\,
\label{XY}
\end{equation}
are also $\cal I$-invariant variables on the sphere.
Thus, a convenient inversion-symmetric family of Hamiltonians is defined by 
\begin{align}
\hat {\mathrm  H}_{\rm gen}
&=
-2(1+|z|^2)^2\,\partial_z\partial_{\bar z}\nonumber\\
&+\sum_{n\ge1} \frac{g_n}{2}\,X_n(|z|) 
+U_{\rm aniso}(X,Y)\,,
\la{Hg}
\end{align}
where  $U_{\rm aniso}$ is any real function of
$X,Y$. The term proportional to $g$ is the Higgs harmonic term, the $g_n$ give higher
order radial anharmonic corrections, and $U_{\rm aniso}$ breaks the axial $U(1)$ symmetry
generated by the shape-space angular momentum operator
\[
\hat J=z\partial_z-\bar z\partial_{\bar z}\,.
\]
In particular, a linear term
\begin{equation}
U_{\rm aniso}(X,Y) \simeq -\epsilon_x X-\epsilon_y Y  \,.
\label{tilt}
\end{equation}
tilts the Higgs well in the direction $\boldsymbol \epsilon = (\epsilon_x, \epsilon_y)$ displacing 
the minimum away from the equilateral pole $z=0$.
Because the inversion symmetry (\ref{inv}) is retained, any minimum at $z_\ast$ is
accompanied by a mirror minimum at $1/\bar z_\ast$.
The exact $(n,m)$-lattice description used in the main text is recovered in the isotropic 
Higgs subfamily
\begin{equation}
g_n=0\quad (n\ge2),
\qquad
U_{\rm aniso}=0  \,,
\la{isof}
\end{equation}
but for generic $g_n$ and $U_{\rm aniso}$, the model \eqref{Hg} is no longer exactly
solvable. 

\subsection{Exact isotropic Higgs oscillator with general \texorpdfstring{$g$}{g}}

In the isotropic harmonic case, thus setting \eqref{isof}, the Hamiltonian reads as
\[
\hat {\mathrm H}_g
=
-2(1+|z|^2 )^2\,\partial_z\partial_{\bar z}
+\frac{g}{2}\,\frac{|z|^2}{(1-|z|^2)^2}\,.
\]
Introducing the spherical variables
\[
z=\tan\frac{\Theta}{2}\,e^{i\phi}
\quad\text{and}\quad
0\le \Theta<\frac{\pi}{2}\,,
\]
this Hamiltonian becomes
\[
\hat {\mathrm H}_g
=
-2\left[
\frac1{\sin\Theta}\partial_\Theta(\sin\Theta\,\partial_\Theta)
+\frac1{\sin^2\Theta}\partial_\phi^2
\right]
+\frac{g}{8}\tan^2\Theta \,,
\]
with $\Theta=\pi/2$ the equatorial Higgs barrier.
Separating variables as
\[
\psi(\Theta,\phi)=e^{im\phi}R(\Theta),
\quad\text{with}\quad m\in\mathbb Z\,,
\]
and defining
\[
\alpha_g:=\sqrt{g+4}\,,
\]
one may write
\[
R(\Theta)
=
(\sin\Theta)^{|m|}
(\cos\Theta)^{\frac12+\frac{\alpha_g}{4}}
F(\cos 2\Theta)\,.
\]
The equation in $\Theta$ reduces to the Jacobi equation, and regularity on 
the hemisphere selects the Jacobi polynomials
\[
F(x)=P_{l}^{\left(|m|,\frac{\alpha_g}{4}\right)}(x),
\qquad
l=0,1,2,\dots
\]
Equivalently, with 
\[
n=|m|+2l,
\qquad
n=0,1,2,\dots
\]
and 
\[
|m|\le n,
\quad\text{with}\quad\,,
n-m\in2\mathbb Z 
\]
the normalized eigenfunctions may be written as \cite{Higgs-1979,Leemon-1979}
\begin{align}
\psi_{nm}(z,\bar z)
=
\mathcal N_{nm}\,
&e^{im\phi}
(\sin\Theta)^{|m|}
(\cos\Theta)^{\frac12+\frac{\alpha_g}{4}} \times
\nonumber\\
&\times P_{\frac{n-|m|}{2}}^{\left(|m|,\frac{\alpha_g}{4}\right)}(\cos 2\Theta)
\end{align}
with spectrum
\[
E_n^{(g)}
=
2(n+1)^2+\alpha_g\,(n+1)\,,
\]
so that  the energy depends only on $n$, not on $m$, as expected for the rotationally
symmetric Higgs oscillator.

Near the equilateral point $z=0$, one has
\[
\hat {\mathrm H}_g
=
-2\,\partial_z\partial_{\bar z}
+\frac{g}{2}\,z\bar z
+O\!\left(|z|^4,\;|z|^2\partial_z\partial_{\bar z}\right)\,,
\]
so that  the tangent-plane limit is the standard two-dimensional harmonic oscillator with local frequency $\omega_{\rm loc}=\sqrt g$.

\subsection{Support parabola and generic two-state encodings}

For the isotropic Higgs oscillator, following the main text any relatively stationary state expanded in the basis
$\{\ket{n,m}\}$ satisfies
\[
E_n^{(g)}=\lambda_0+\lambda_J m\,.
\]
Since $E_n^{(g)}$ is quadratic in $n$, this defines a parabola on the admissible
$(n,m)$-lattice,
\[
2(n+1)^2+\alpha_g(n+1)=\lambda_0+\lambda_J m\,.
\]

Let $(n_a,m_a)$ and $(n_b,m_b)$ be two admissible lattice points with $m_a\neq m_b$.
The support parabola through them is fixed by
\[
\lambda_J=
\frac{E_{n_a}^{(g)}-E_{n_b}^{(g)}}{m_a-m_b},
\qquad
\lambda_0=
\frac{m_aE_{n_b}^{(g)}-m_bE_{n_a}^{(g)}}{m_a-m_b}.
\]

Now suppose that a third lattice point $(n,m)$ also lies on the same support parabola.
Then
\[
(m_a-m_b)E_n^{(g)}
-m_aE_{n_b}^{(g)}
+m_bE_{n_a}^{(g)}
-\bigl(E_{n_a}^{(g)}-E_{n_b}^{(g)}\bigr)m
=0
\]
Substituting 
\[
E_n^{(g)}=2(n+1)^2+\alpha_g(n+1)
\] 
one finds a rational part and an
$\alpha_g$-part. For generic $g$, 
\[
\alpha_g=\sqrt{g+4}\notin\mathbb Q\,,
\] 
so the two parts
must vanish separately. The $\alpha_g$-part gives
\begin{equation}
m=
\frac{(m_a-m_b)n+n_am_b-n_bm_a}{n_a-n_b}\,,
\la{line-g}
\end{equation}
which is the line through the two chosen lattice points, while the rational part reduces to
\[
2(n-n_a)(n-n_b)=0\,.
\]
Therefore $n=n_a$ or $n=n_b$, and then (\ref{line-g})  implies $m=m_a$ or $m=m_b$.
Hence, for $\lambda_J\neq0$ and generic $g$, the support parabola through two admissible
lattice points contains no further admissible lattice points. In particular, relative stationary
qubit encodings are generically two-state encodings in the isotropic Higgs theory.

The exceptional set $\alpha_g\in\mathbb Q$, i.e. $g=q^2-4$ with $q\in\mathbb Q$, is
countable; outside this set the above two-point argument holds exactly.

\subsection{Specialization used in the main text}

The Hamiltonian used in the main text reads
\begin{equation}
\hat {\mathrm H}
=
-2(1+z\bar z)^2\,\partial_z\partial_{\bar z}
+\frac12\,\frac{z\bar z}{(1-z\bar z)^2} ,
\la{our-H}
\end{equation}
with the spectrum
\begin{equation}
E_n
=
2(n+1)^2+\sqrt5\,(n+1)\,.
\label{our-spectrum}
\end{equation}
The choice $g=1$ of the main text is distinguished because (\ref{our-H})  becomes
\begin{align}
\hat {\mathrm H}
&=
-2\,\partial_z\partial_{\bar z}
+\frac12\,z\bar z
+O\!\left(|z|^4,\;|z|^2\partial_z\partial_{\bar z}\right)   \nonumber\\
&= - \frac{1}{2} (\partial_x^2 + \partial_y^2) + \frac{1}{2} (x^2 + y^2) + \dots\,,
\label{our-flat}
\end{align}
so the tangent-plane limit about the equilateral point $z=0$ is exactly the standard
two-dimensional harmonic oscillator in oscillator units. Thus the Hamiltonian used in the main text
is the unique member of the isotropic family \eqref{Hg} whose local flat-space limit
has unit oscillator frequency and harmonic coefficient $1/2$.

\subsection{The geometric phase}

Here we derive the geometric phase presented in the main text.
For this we recall
the minimum-energy relative-stationary state 
\begin{align}\label{Supp:miniE}
 & \ket{\psi_{k\ell}(t)}  =
\\
&
\, e^{-iE_k t} \! \left[  \sqrt{p^{(k)}_0}\,\ket{k,k} +
e^{ i(\varphi_k- \Delta_k t) } \sqrt{p^{(k)}_1}\,\ket{k+1,k+1} \right].\nonumber
\end{align}
After one density shape-space
rotation with period $T_{\rm rot}^{(k)}= {2\pi}/{\Delta_k}$, the state becomes
\[
|\psi_{k\ell}(T_{\rm rot}^{(k)})\rangle =
e^{-iE_k T_{\rm rot}^{(k)}}\ket{\psi_{k\ell}(0)} \,,
\]
with total cyclic phase
\[
\beta_{\rm tot} = -E_kT_{\rm rot}^{(k)} =
-2\pi\frac{E_k}{\Delta_k} \qquad
({\rm mod}\ 2\pi)\,.
\]
The dynamical phase accumulated over this cycle is 
\[
\beta_{\rm dyn} =
-\int_0^{T_{\rm rot}^{(k)}}dt\,
\bra{\psi_{k\ell}(t)}\hat{\mathrm H}\ket{\psi_{k\ell}(t)} 
\]
Since the probabilities of the two support states $\ket{k,k}$ and $\ket{k+1,k+1}$ 
are time independent,
\[
\bra{\psi_{k\ell}(t)}\hat{\mathrm H}\ket{\psi_{k\ell}(t)}
= (1-x)E_k+xE_{k+1} = E_k+x\Delta_k \,.
\]
Therefore
\[
\beta_{\rm dyn} = -\left(E_k+x\Delta_k\right)
\frac{2\pi}{\Delta_k} = -2\pi \left(
\frac{E_k}{\Delta_k}+x \right)\,.
\]
The geometric phase is then obtained by removing  the dynamical phase from the cyclic phase,
\[
\beta_{\rm geom} = \beta_{\rm tot}-\beta_{\rm dyn}
= 2\pi x = 2\pi(\ell-k) \qquad ({\rm mod}\ 2\pi)\,.
\]
Using the Bloch-sphere parametrization
\[
x=\ell-k=\sin^2\frac{\theta_k}{2},
\]
this becomes
\[
\beta_{\rm geom} = \pi(1-\cos\theta_k)
\qquad ({\rm mod}\ 2\pi)\,.
\]
where the sign follows from the convention
$\varphi_k(t)=\varphi_k-\Delta_k t$.  If the same latitude is traversed in
the opposite direction, the geometric phase changes sign.

For completeness we also note why the observable density period is independent
of $\ell$:  Since the  two states in \eqref{Supp:miniE} differ by one unit of
shape angular momentum,  the time-dependent part of
the density is the interference term
\[
\rho_{\rm int} \propto 2\sqrt{x(1-x)}
\cos\!\left[\phi+\varphi_k-\Delta_k t\right],
\]
where $\phi$ is the azimuthal angle on the shape sphere.
The frequency of
this rotating pattern is $\Delta_k$, while $\ell$ only controls the
visibility $2\sqrt{x(1-x)}$, which vanishes at the endpoints
$\ell=k$ and $\ell=k+1$.

\vspace{0.4cm}

\subsection{Animations}

Animations can be found as ancillary files, or at the following link 
\normalsize{\url{http://www.idpoisson.fr/garaud/research/geometric-qubit.html}}.

\begin{itemize}
\item \textbf{Video 1:} \hfill{\footnotesize{\tt Kendall-shape-space-orbit.[avi/mp4]}}\\  This animation  exemplifies how different points on Kendall  two-sphere correspond to different triangle shapes, depicted  in Figure 1 of main text. It also shows how a trimer deformation cycle in real space maps to a close orbit around north pole of Kendall's shape sphere. The yellow equilateral triangles below the trimers correspond to the north pole on Kendall's sphere around which the closed orbit rotate. 

\item \textbf{Video 2:} \hfill{\footnotesize{\tt Wavefunction-evolution.[avi/mp4]}}\\
This animation shows the time evolution of a relative stationary state 
depicted  in Figure 2 of main text.
\end{itemize}

%

\end{document}